\documentclass[%
reprint,
 amsmath,amssymb,
 aps,
prb,
]{revtex4-1}

\usepackage{graphicx}
\usepackage{dcolumn}
\usepackage{bm}
\usepackage{appendix}
\usepackage{color}
\usepackage{braket}

\begin{document}
\newcommand{\laplacian}{\nabla^2}

\preprint{APS/123-QED}

\title{Soft-mode and Anderson-like localization in two-phase disordered media}

\author{Tingtao Zhou}
\thanks{email: edmondztt@gmail.com\\
Current address: Department of Mechanical and Civil Engineering, Caltech}
\affiliation{Department of Physics, Massachusetts Institute of Technology}
\author{Dimitrios Fraggedakis}
\thanks{aka dfrag - 
email: dimfraged@gmail.com}
\affiliation{Department of Chemical Engineering, Massachusetts Institute of Technology}%
\author{Fan Wang}
\thanks{email: fwangmit@gmail.com}
\affiliation{Department of Mathematics, Massachusetts Institute of Technology}

\date{\today}

\begin{abstract}
Wave localization is ubiquitous in disordered media -- from amorphous materials, where soft-mode localization is closely related to materials failure, to semi-conductors, where Anderson localization leads to metal-insulator transition. Our main understanding, though, is based on discrete models. Here, we provide a continuum perspective on the wave localization in two-phase disordered elastic media by studying the scalar wave equation with heterogeneous modulus and/or density. At low frequencies, soft modes arise as a result of disordered elastic modulus, which can also be predicted by the localization landscape. At high frequencies, Anderson-like localization occurs due to disorder either in density or modulus. For the latter case, we demonstrate how the vibrational dynamics changes from plane waves to diffusons with increasing frequency. Finally, we discuss the implications of our findings on the design of architected soft materials.
\end{abstract}

\maketitle

\section{Introduction\label{sec:intro}}

Disordered media have enormous potential in engineering applications, such as random lasers for speckle-free imaging~\cite{cao2003lasing,redding2012speckle}, amorphous semiconductors~\cite{davis1970conduction,anderson1975model,madan2012physics}, or multi-functional materials~\cite{kim2020multifunctional} due to their continuous range of tunable properties.
In nature, disorder appears at various length scales and largely affects our understanding on related physical phenomena --
mechanics of cellular cytoskeleton is governed by its network structural disorder~\cite{fletcher2010cell,chaubet2020dynamic};
seismic 
waves propagate through complex composites of soil, water and rocks~\cite{kennett2009seismic,sato2012seismic}; the density contrast between multiple phases of molecular clouds influence the formation of stars and planets~\cite{huang2013coagulation,zhou2015imf}.
Not surprisingly, an increasing research interest is devoted to these disordered materials. As a prototype and first order approximation, many of these scenarios can be modeled as two-phase media with contrasting components.

Since the seminal paper of Anderson~\cite{anderson1958absence}, the connection between wave localization and disorder has been a fascinating research topic and a best example of his philosophical statement ``more is different''~\cite{anderson1972more}. For discrete systems, a powerful tool to study localization is the random matrix theory~\cite{beenakker1997random} and its more recent version free probability theory~\cite{chen2012error,movassagh2017eigenpairs,welborn2013densities}. The original observations were that all single particle states in one-dimensional potentials with arbitrarily weak disorder decay exponentially~\cite{anderson1958absence}, and a threshold amount of disorder in three-dimensions for all states to be localized~\cite{abrahams1979scaling}. Two dimensions is a marginal case where weak disorder can cause the transport properties of the system to decay logarithmically with the system size~\cite{abrahams1979scaling,lee1981anderson}.

For the continuous counterpart,
a mathematically rigorous distinction between localized and extended modes has been established for a self-adjoint operator supported on an unbounded domain~\cite{pastur1973spectra,kirsch1982spectrum,klopp1995localization,aizenman2015random}.
Due to the Lebesgue decomposition theorem~\cite{hewitt2013real}, the spectrum of this operator can always be written as $\mu = \mu_{pp} + \mu_{sc} + \mu_{ac}$. The pure point spectrum $\mu_{pp}$ is the eigenvalue set with normalizable (localized) eigenfunctions. It is countable, and hence has zero Lebesgue measure, but may be still dense, for example in the original Anderson model with on-site disorder potentials sampled from a continuous distribution~\cite{frohlich1984rigorous}.
The absolutely continuous part $\mu_{ac}$ corresponds to the extended states such as a plane wave, which are common in physics but cannot be represented by normalizable functions in the Hilbert space. 
The singular continuous part $\mu_{sc}$ is more subtle and gives rise to interesting states between localized and extended modes~\cite{cerdeira1991quantum}. In this sense, localized modes such as bound states (soft modes) or Anderson localization is reflected by the pure point spectrum. More generally there are special cases of localization in an infinite domain not induced by disorder but by either symmetry, mismatched boundary conditions between domains or certain  designed potentials, known as bound states in the continuum (BIC)~\cite{von1993merkwurdige,hsu2016bound}, with eigenmodes embedded inside the continuous spectrum. 

Classical bound states appear in elementary quantum mechanics problems, such as the discrete energy levels of a hydrogen atom~\cite{griffiths2018introduction}, as well as in other classical wave problems~\cite{mei2018theory}.
The mechanism for Anderson localization, however, is usually understood as a coherent scattering process in a random potential landscape~\cite{mott1967electrons}, which was first rationalized by the Ioffe-Regel criterion~\cite{ioffe1960non}, and later by the Thouless criterion~\cite{edwards1972numerical} for the scaling of Thouless conductance.
In contrast to the classical bound state mechanism~\cite{kittel1996introduction}, the total energy of the single particle is not necessarily lower than the maximum of the random potential~\cite{aspect2009anderson}, similar to BIC~\cite{von1993merkwurdige}. Beyond the analysis for single particle Hamiltonians, recent studies have also shown localization in interacting many-body quantum systems~\cite{aizenman2009localization}, while the phase diagram of many-body localization remains under study~\cite{potter2015universal}.

Here, we are interested in the behavior of the continuous differential operator that describes the scalar wave propagation in two-dimensional finite domains. 
Scalar wave localization has been considered analytically for two component composites~\cite{sheng1986scalar}, where a minimum impedance contrast was found by scaling argument to cause localization in an infinite domain. The above mentioned mathematically rigorous concept of localization does not apply to problems defined on finite domains. However, it is still of interest in engineering applications to compare different states and quantify how extended/localized they are inside the finite domain by proper order parameters -- for example the inverse participation ratio (IPR) that is often used in studying disordered systems~\cite{wegner1980inverse,murphy2011generalized}. 

In this work, we examine the statistics of the eigenvalue spectrum and inverse participation ratio for the continuous scalar wave equation in finite two-dimensional domains of length $L$. In particular, we consider a two-phase medium with densities and moduli $\rho_1,G_1$ and $\rho_2,G_2$, respectively, with correlation length $l_c$. For disorder in $G$, we identify the values of $G_1/G_2$ for which soft mode localization is present, and discuss the effect of correlation length. For disorder in $\rho$, we find the existence of high frequency modes for $\rho_1/\rho_2<O(10^{-2})$, referred to as Anderson-like modes.
Based on these results, we show the observation of cross-over from propagation to diffusion in a two-phase medium as the frequency of the externally applied force increases, reminiscent of the Ioffe-Regel frequency.
Our goal is to present a phenomenological description on the trend of wave localization for disordered two-phase media, clarify connections/distinctions of different localization mechanisms, and motivate further experimental and engineering work on architectured materials.

\section{Theory of linear elastic waves\label{sec:theory}}

\begin{figure}
    \centering
    \includegraphics[width=\linewidth]{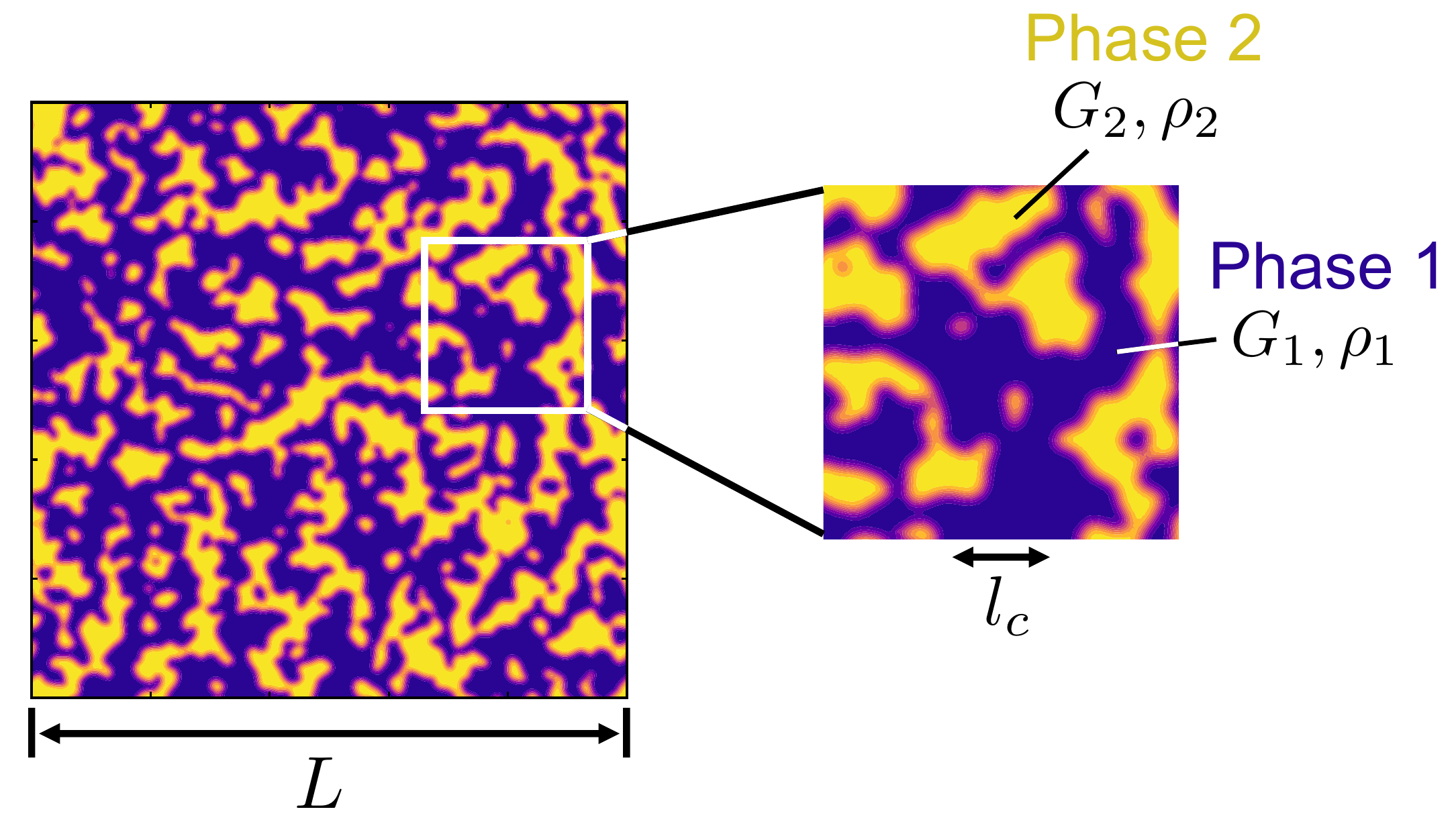}
    \caption{
    Example illustration of a two-phase disordered medium, with properties $\rho_1,G_1$ (purple) and $\rho_2,G_2$ (yellow), respectively. The system has finite length $L$ and the spatial distribution of the two phases is characterized by the correlation length $l_c$. Without loss of generality, we assume the physical properties of phase 2 to have larger magnitude than those of phase 1.
    }
    \label{fig:model_system}
\end{figure}

The propagation of a linear elastic wave is commonly described with the following set of equations~\cite{gurtin2010mechanics}
\begin{subequations}
\begin{equation}
    \rho \ddot{\bm{u}} = \nabla \cdot \bm{\sigma} + \bm{b}
\end{equation}
\begin{equation}
    \bm{\sigma} = \mathbb{C} \bm{\epsilon}
\end{equation}
\begin{equation}
    \bm{\epsilon} = \frac{1}{2} \left(
    \nabla \bm{u} + (\nabla\bm{u})^T
    \right)
\end{equation}
\label{eqn:elastic-wave-general}
\end{subequations}
where $\bm{u}$ is the displacement field, $\bm{\sigma}$ the stress tensor, $\bm{\epsilon}$ the strain tensor, $\bm{b}$ the body force per unit volume, $\rho$ the density, and $\mathbb{C}$ the stiffness tensor of the material. For a single Fourier mode of frequency $\omega$ and assuming negligible body forces and isotropic elasticity, Eq.~\ref{eqn:elastic-wave-general}(a) is known as the Cauchy-Navier equation
\begin{equation}
\label{eqn:cauchy-navier}
\begin{split}
    - \omega^2 \rho \bm{u} = &
\nabla\cdot \bm{u} \nabla\lambda + 
\left[\nabla\bm{u} +
 (\nabla\bm{u})^T
\right] \cdot \nabla \mu\\
&+ \left(\lambda + \mu\right) \nabla(\nabla\cdot\bm{u}) + \mu \nabla^2 \bm{u}
\end{split}
\end{equation}
where $\lambda\left(\mathbf{x}\right)$ and $\mu\left(\mathbf{x}\right)$ are the spatially dependent Lam\'{e} coefficients.

When only density is described by a heterogeneous field, the Lam\'{e} coefficients are constant and it is straightforward to adopt the Helmholtz-Hodge decomposition
\begin{equation}\label{eqn:hlmtz_hodge}
\bm{u} = \nabla \phi + \nabla\times\bm{A} = \bm{u}_l + \bm{u}_t
\end{equation}
to simplify the equations. Substituting Eq.~\ref{eqn:hlmtz_hodge} in Eq.~\ref{eqn:cauchy-navier}, we can separate the Cauchy-Navier equation in three polarizations, one longitudinal $u_l$ and two transversal $\bm{u}_t$ as
\begin{equation}
\begin{split}
    -\omega^2 {u}_l & = \frac{\lambda+2\mu}{\rho} \nabla^2 {u}_l \\
    -\omega^2 \bm{u}_t & = \frac{\mu}{\rho} \nabla^2 \bm{u}_t
\end{split}
\end{equation} 

In the case of disordered stiffness tensor, the Lam\'{e} coefficients are also heterogeneous, and Eq.~\ref{eqn:cauchy-navier} cannot be further simplified. As a result, the separation of the generated displacement field in different polarizations is not trivial. 

Herein, we want to analyze the effects of disorder in a two-phase medium on the vibrational modes of a two-dimensional finite domain. To do so, we consider the scalar elastic vibration equation
\begin{equation}
    -\rho\left(\mathbf{x}\right)\omega^2 u =  \nabla\cdot \left(G\left(\mathbf{x}\right) \nabla u\right)
\label{eqn:simplified-wave-1d}
\end{equation}
where $G$ is the elastic modulus. Fig.~\ref{fig:model_system} illustrates the two-dimensional finite size system of length $L$, where the two phases are shown with purple (phase 1) and yellow (phase 2), respectively. The properties of phase 1 are $\rho_1,G_1$ and those of phase 2 are $\rho_2,G_2$. We always assume the properties of phase 2 to be larger in magnitude compared to phase 1. Also, the spatial distribution of the two-phase medium has correlation length $l_c$. The characteristic length, density and modulus are chosen as $L$, $\rho_2$ and $G_2$, so that the frequency can be rescaled as $\omega^2\rightarrow\omega^2G_2/L^2\rho_2$. Thus, we arrive at the following dimensionless form
\begin{equation}
    -\Tilde{\rho}\omega^2 u = \nabla\cdot \left(\Tilde{G}\nabla u\right)
\label{eqn:simplified-wave-1d}
\end{equation}
where $\Tilde{G}=G/G_2$ and $\Tilde{\rho}=\rho/\rho_2$. At the boundaries of the domain we consider Dirichlet conditions $u=0$. We analyze the spectrum of Eq.~\ref{eqn:simplified-wave-1d} by solving a generalized eigenvalue problem. We discretize the differential operator using the finite element method~\cite{fraggedakis2017discretization}, where the details on both the numerical implementation and the eigenvalue problem are given in the Appendix of the paper. In all cases, we analyze the data collected from an ensemble of 100 realizations with volume fraction $\phi=0.5$. In the remaining of the text, we always refer to the dimensionless form Eq.~\ref{eqn:simplified-wave-1d} and drop the tilde symbols.

\section{Results\label{sec:results}}

\subsection{Density disorder and Anderson-like localization\label{sec:anderson-like modes}}

\begin{figure*}[!ht]
    \centering
    \includegraphics[width=0.6\linewidth]{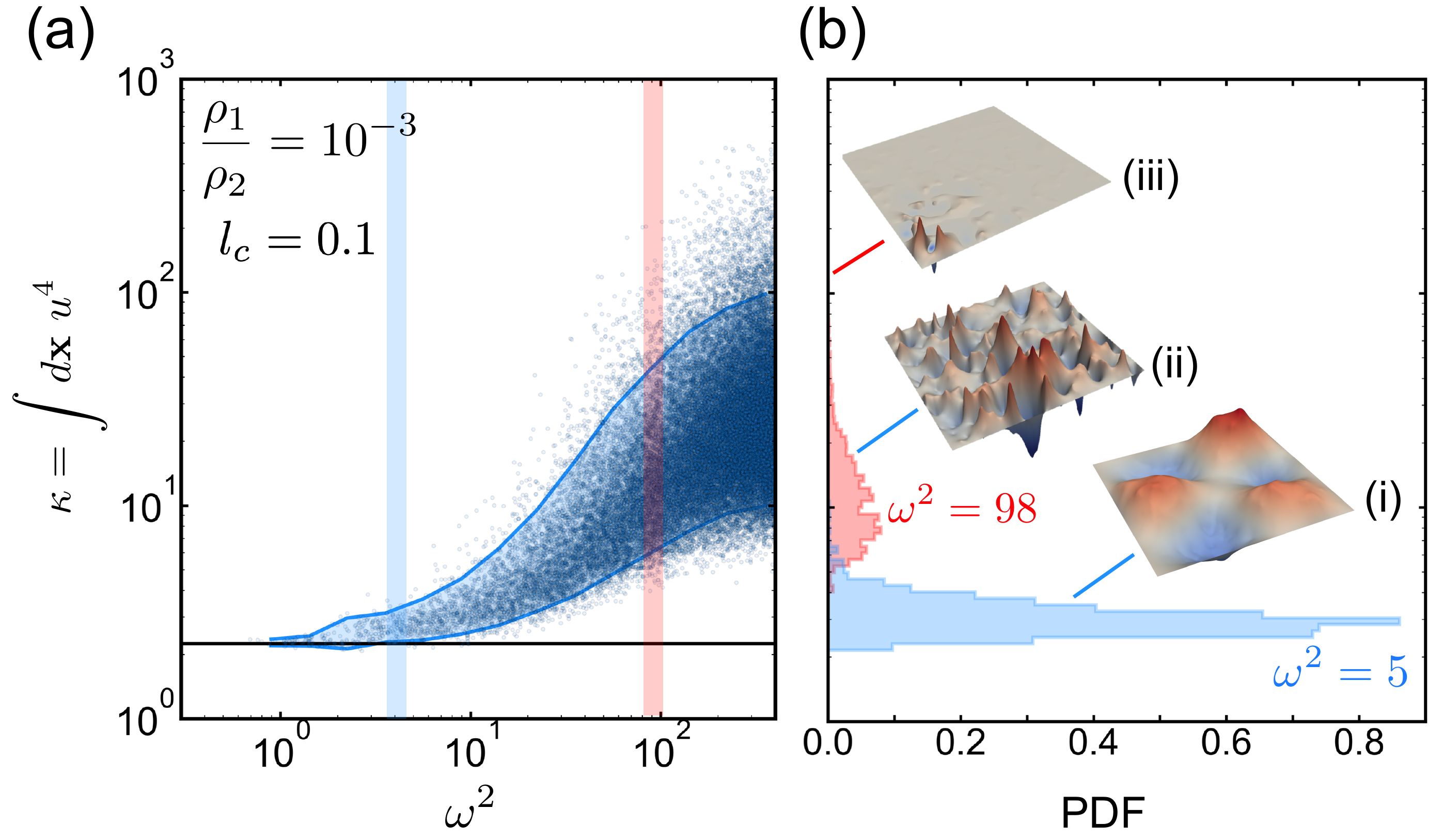}
    \caption{(a) The inverse participation ratio $\kappa=\int d\mathbf{x} u^4$ calculated for an ensemble of 100 realizations of disordered $\rho$, with the lowest 900 eigenvalues from each realization. The density ratio is $\rho_1/\rho_2=10^{-3}$ and the correlation length $l_c=0.02$. The horizontal black line shows the lowest possible value of $\kappa=2.25$ that appears for eigenmodes in a homogeneous domain $\rho\equiv$ constant, which are of the form $u(x,y)=A_{n_1,n_2} \sin(n_1 x)\sin(n_2 y)$.
    The shaded area (boundary marked with blue lines) is the 20-80\% quantile range for the $\kappa$ distribution at each $\omega^2$.
    (b) Probability density function of $\kappa$ for $\omega^2=5$ (blue shaded line in (a)) and $\omega^2=98$ (red shaded line in (a)). As inset, we include the wave profiles for (i) $\kappa\simeq3$, (ii) $\kappa\simeq20$ and (iii) $\kappa\simeq100$. The highest value of $\kappa$ corresponds to an Anderson-like localized mode.
    }
    \label{fig:localization-index-rho}
\end{figure*}

We first examine the role of disordered density profile on the eigenvalue and eigenvector spectrum of Eq.~\ref{eqn:simplified-wave-1d}. In this case, we keep ${G}=1$ for all the realizations studied.

Figure~\ref{fig:localization-index-rho}(a) shows the statistics of IPR with $\rho_1/\rho_2=10^{-3}$ and correlation length $l_c=0.1$. The IPR is defined as $\kappa=\int d\mathbf{x}\,\,u^4$ with normalization for the eigenvectors $u(\mathbf{x})$ being $\int d\mathbf{x}\,\,u^2=1$. Therefore, in the case of no disorder, all the standing waves in a homogeneous domain result in $\kappa=2.25$. The blue lines correspond to the $20-80\%$ quantile range, and the black line indicates the lower bound for the IPR. Larger IPR indicates higher degree of localization.

For low frequency modes, $\omega^2 \sim 1$, all realizations produce standing waves that span the entire domain. As a result, density disorder does not affect the low frequency spectrum. On the contrary, as the frequency increases and the wavelength approaches the correlation length, $\omega\sim l_c^{-1}$, the eigenvectors attain a more localized profile than in the low frequency regime. In this case, IPR starts increasing to values much larger than the baseline of $\kappa=2.25$, as shown in Fig.~\ref{fig:localization-index-rho}(a). For $\omega^2>10^2$, we observe IPR to span approximately two orders of magnitude, $\kappa\sim O(1)-O(10^2)$. The statistical sampling demonstrates that for large enough frequencies there are going to be extended, but fairly non-uniform, standing waves, as well as highly localized ones. To better understand this result, we focus on the statistics of $\kappa$ around specific values. 

Figure~\ref{fig:localization-index-rho}(b) shows the distribution of IPR for a low frequency $\omega^2= 5$ (blue) and a relatively higher one $\omega^2 = 98$ (red). When the frequency is low, the distribution has a sharp peak at low values of $\kappa$ that corresponds to extended standing waves, as in Fig.~\ref{fig:localization-index-rho}(b)-(i). In the high frequency case, the distribution of $\kappa$ is much more extended spanning two orders of magnitude, thus the small-scale details of $\rho$ are expected to largely affect the solution profiles. We examine two realizations which result in $\kappa\simeq20$, Fig.~\ref{fig:localization-index-rho}(b)-(ii), and $\kappa\simeq100$, Fig.~\ref{fig:localization-index-rho}(b)-(iii). For moderate IPR, the eigenfunction extends throughout the entire domain, however it is highly heterogeneous. For large IPR, though, the solution profile is qualitatively different as it is highly localized nearby the boundary of the domain. Since these localized modes share the characteristic of appearing at high frequencies reminiscent of the original Anderson localization in infinite domains, we refer to them as Anderson-like localization. 

\subsection{Modulus disorder and soft mode localization\label{sec:soft modes}}

\begin{figure*}[!ht]
    \centering
    \includegraphics[width=\linewidth]{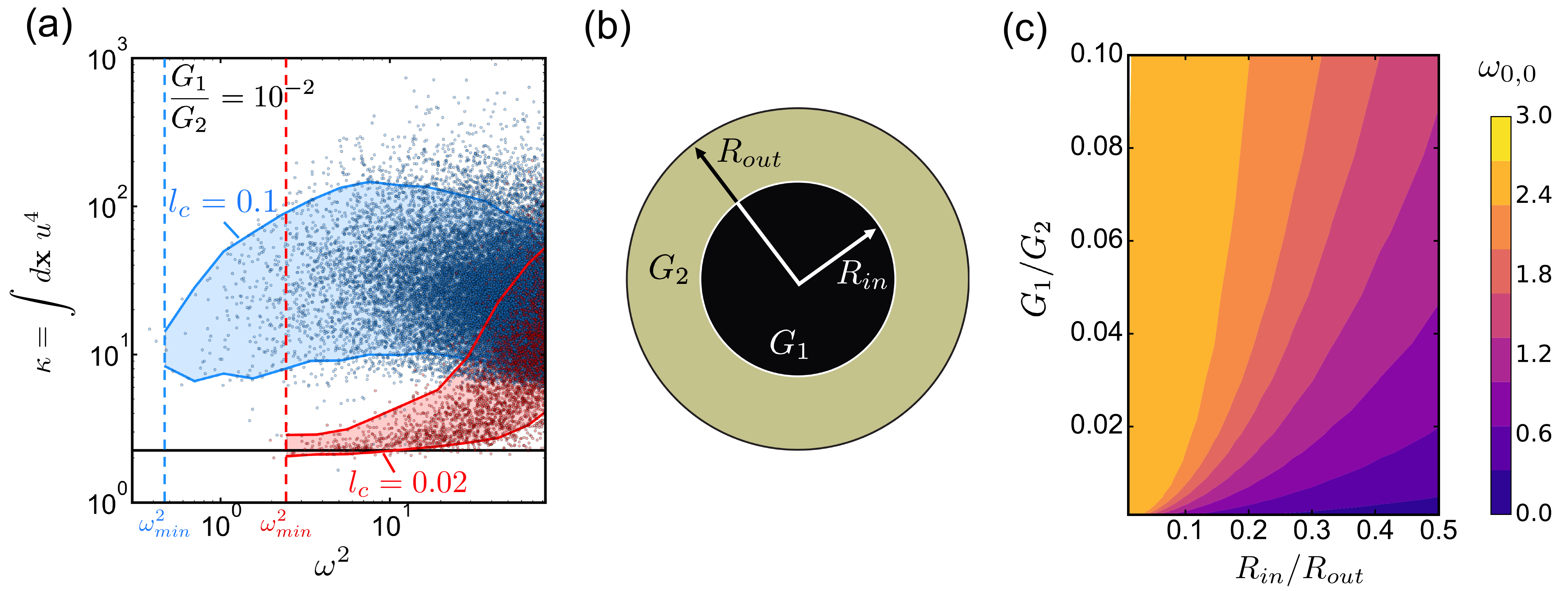}
    \caption{(a) The inverse participation ratio $\kappa=\int d\mathbf{x} u^4$ calculated for an ensemble of 100 realizations of disordered $G$, with the lowest 900 eigenvalues from each realization. The modulus ratio is $G_1/G_2=10^{-2}$. With red symbols/shade we show the results for $l_c=0.1$, and with blue symbols/shade we depict the statistics for $l_c=0.1$. The description of the horizontal black line is given in Fig.~\ref{fig:localization-index-rho}.
    The shaded areas are the 20-80\% quantile range for the $\kappa$ distribution at each $\omega^2$. We find that with increasing $l_c$, low frequency modes becomes localized which are also known as soft modes.
    (b) Schematic for the single bound state analysis testing the effect of correlation length. We perform the analysis for a composite shaped as concentric circles with inner and outer radii and scalar modulus $R_{in}$, $R_{out}$, $G_1$, $G_2$, respectively. The effects of correlation length are given by changing the ratio $R_{in}/R_{out}\sim l_c$.
    (c) Contour plot for the ground state eigenvalue $\omega_{0,0}$ as a function of $G_1/G_2$ and $R_{in}/R_{out}$. For fixed scalar modulus ratio, decreasing $l_c\sim R_{in}/R_{out}$ corresponds to increasing $\omega_{0,0}$, which is consistent with the trend shown by the two vertical dashed lines marked on (a).
}
    \label{fig:localization-index-G}
\end{figure*}

In this section, we examine how the heterogeneous modulus $G$ affects both the eigenvalue spectrum and the spatial distribution of the eigenmodes with the density set to be homogeneous $\rho=1$. 

First, the spatial dependence of $G$ allows us to rewrite Eq.~\ref{eqn:simplified-wave-1d} in the following form
\begin{equation}\label{eqn:elastic_potential}
- \frac{\omega^2}{G} u = \nabla^2 u + \frac{\nabla G}{G} \nabla u
\end{equation}
which is similar to the Schr\"{o}dinger equation with random potential $V=\nabla G/G$, except that the ``energy'' $E=\omega^2 / G$ is spatially dependent. As we will shortly see, the correspondence between the Schr\"{o}dinger equation and the scalar elastic vibration equation helps to explain both  low and high frequency localized modes from this operator. 

Figure~\ref{fig:localization-index-G}(a) demonstrates the statistics of IPR for two different correlations lengths, $l_c=0.02$ (red) and $l_c=0.1$ (blue). The two cases demonstrate several qualitative differences. First, the results for $l_c=0.02$ show similar trend with those obtained with density disorder, where at low frequencies the waves are extended ($\kappa\simeq2.25$) and at high frequencies Anderson-like modes emerge. The IPR statistics for $l_c=0.1$, however, does not resemble the previous cases, as even the lowest frequency modes have $\kappa$ around $10$. The low frequency modes with large $\kappa$ correspond to soft modes~\cite{tanguy2010vibrational}, which are localized eigenvectors of Eq.~\ref{eqn:simplified-wave-1d}. For large frequencies, the soft modes are transformed into Anderson-like modes, like in the cases of $l_c=0.02$ and density disorder. 

The difference between lowest eigenvalues of the two correlation lengths $l_c$ can be understood by examining the simplified model of Fig.~\ref{fig:localization-index-G}(b). In particular, we consider two concentric cylinders, with the inner modulus $G_1$ and radius $R_{in}$, and the outer modulus $G_2$ and radius $R_{out}$, respectively. In this model, the outer cylinder is the stiffer one. The dimensionless ``correlation'' length $l_c$ in this model corresponds to the ratio $R_{in}/R_{out}$. The analytical solution of the scalar elastic vibration equation for this simplified model is given in the Appendix. In Fig.~\ref{fig:localization-index-G}(c), we present the lowest eigenvalue $\omega_{0,0}$ as a function of the moduli ratio and the correlation length. From the contour plot, it is apparent that for the smallest $G_1/G_2$ and $R_{in}/R_{out}>0.1$, the ground state eigenvalue attains very small values, similar to $l_c=0.1$ where soft modes are present. Decreasing either $R_{in}/R_{out}$ or decreasing the difference between $G_1$ and $G_2$, the ground state frequency increases, recovering the extended mode limit, as in $l_c=0.02$. 

\subsection{From propagating waves to diffusive response\label{sec:diffuson}}

\begin{figure}
    \centering
    \includegraphics[width=\linewidth]{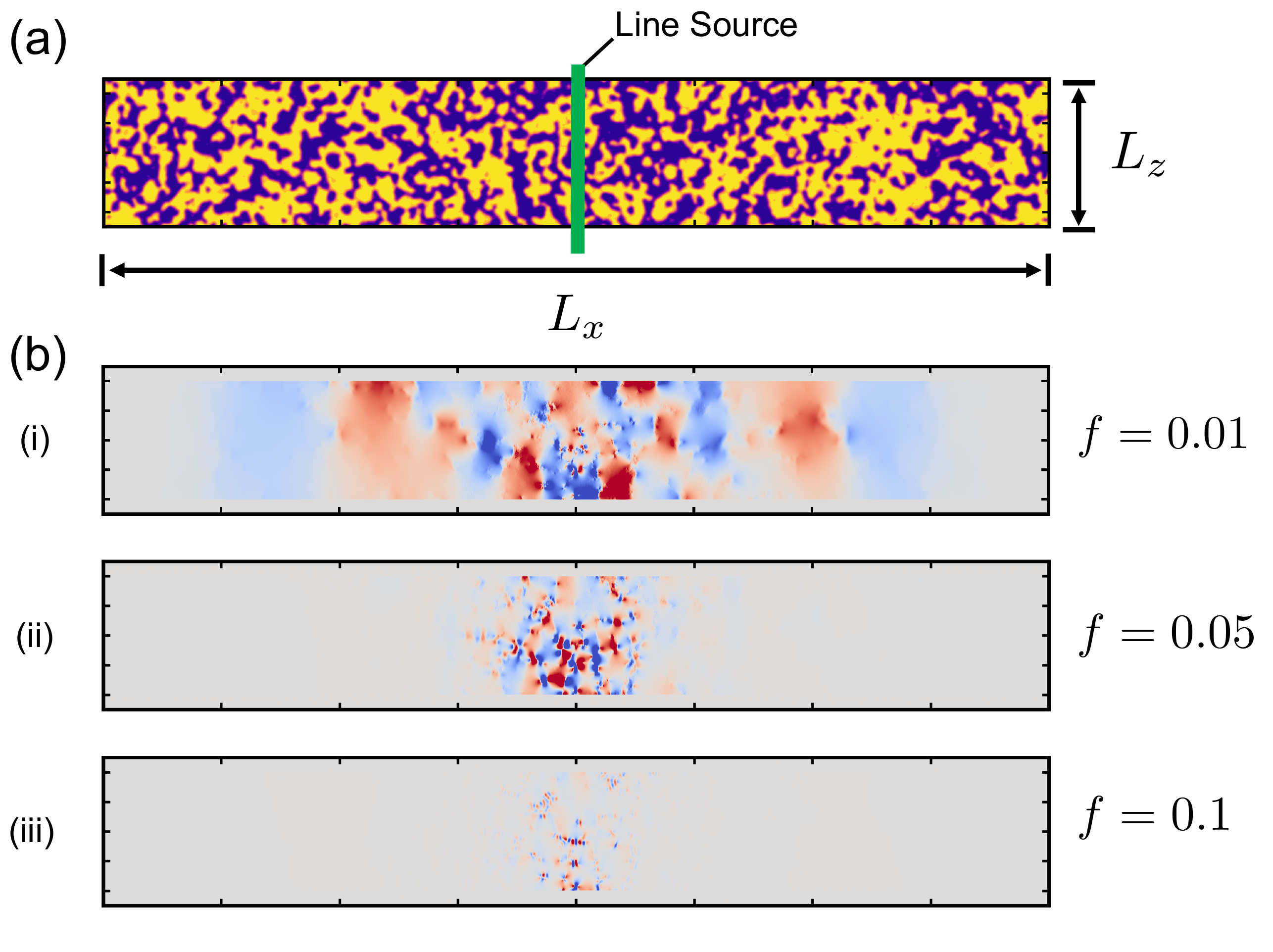}
    \caption{In a domain with disordered $G$, the response to a forced pulse shows a transition from propagating waves to diffusive modes. (a) Snapshot of the system heterogeneity in terms of the modulus $G$. The system has size $L_x=8L$ and $L_z=1L$. The density is constant $\rho=1$, while the ratio is $G_2/G_1=10^2$ and the correlation length equals $l_c=0.02$. A short pulse of the form $u_0(t)=\sin(\Omega t)\delta(x-L_x/2)$ is applied from the line source located in the middle of the domain (green line), with $\Omega=4\pi f \times10^{4} \left( \frac{1~\text{mm}}{L} \sqrt{ \frac{G_2}{1~\text{GPa}} \frac{1~\text{g/cm}^3}{\rho_2} }\right)$~Hz. (b) Snapshots of the propagating wave structures that result at $t=5\times 10^{-3}/\pi$~s from the line source with frequency (i) $f=0.01$ Hz, (ii) $f=0.05$ Hz, (iii) $f=0.1$ Hz.
    }
    \label{fig:diffuson}
\end{figure}

By studying the disorder in both the density and the modulus variables, we reveal the material parameters of a two-phase medium for which the eigenmodes are localized. In this section, we examine the dynamical response of such disordered media to external forcing.
Here, we consider the medium of Fig.~\ref{fig:diffuson}(a), with dimensions $L_x$ and $L_z$, respectively, where $L_x>L_z$. For demonstration, we consider heterogeneous elastic modulus $G$ only, with $G_1/G_2=10^{-2}$ and correlation length $l_c=0.02$. Based on the analysis of Fig.~\ref{fig:localization-index-G}(a), we expect the existence of Anderson-like eigenmodes at high frequencies. At $x=L_x/2$, we place a line source which produces a pulse of the form $u_0\left(t\right)=\sin(\Omega t)\delta(x-L_x/2)$, where $\Omega$ is the excitation frequency. In this part we provide real units quantities to demonstrate the principles for experimental design. On the upper and lower sides of the domain we set a stress-free boundary condition, while on the left and right sides we assume open boundaries. The simulations were conducted using the finite-difference-time-domain method in MEEP package~\cite{taflove2013advances,oskooi2010meep}. More details on the numerical implementation are given in the Appendix.

Figures~\ref{fig:diffuson}(b) (i)-(iii) show three cases with different imposed frequencies, i.e. (i) $f = 0.01$ Hz, (ii) $f = 0.05$ Hz, and (iii) $f = 0.1$ Hz. For low enough frequencies, $f<0.01$ Hz, modulated plane wave propagation is observed, Fig.~\ref{fig:diffuson}(b)-(i). However, once the wavelength of the imposed pulse is comparable to the correlation length $l_c$, the transmission of the wave is weakened, Fig.~\ref{fig:diffuson}(b)-(ii). Increasing the frequency of the pulse to values larger than $f>0.1$ Hz, the wave gets localized around the source. This phenomenon, that with increasing frequency  waves become localized, was also observed in molecular models of amorphous solids~\cite{beltukov2018propagative,seyf2016method}.

\section{Discussion}

\subsection{The different faces of localization phenomena\label{sec:nature of localization-EM}}
\begin{figure*}
    \centering
    \includegraphics[width=0.8\linewidth]{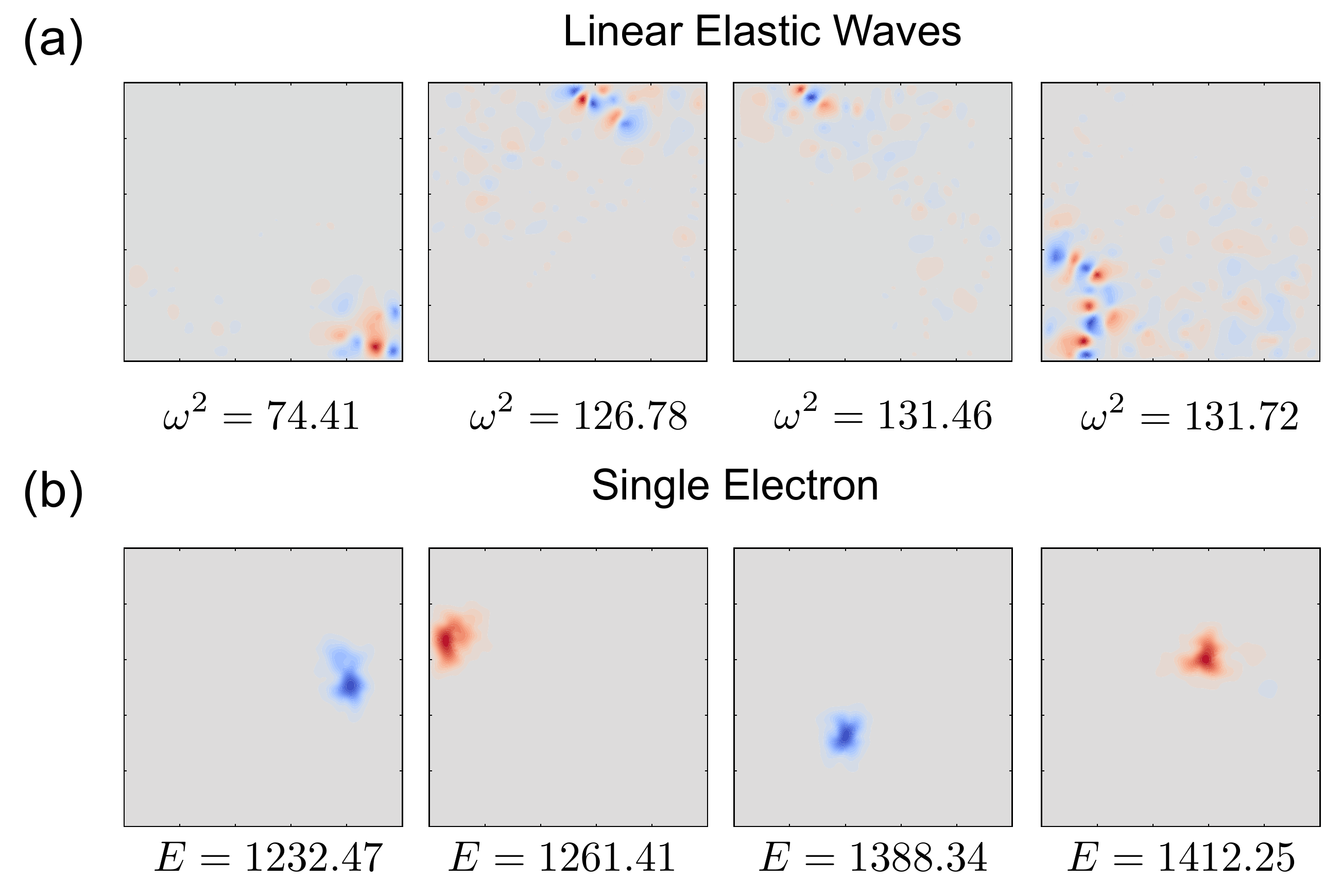}
    \caption{
    Comparison of localized modes from scalar wave equation and single particle Schr\"{o}dinger equation. The disorder is introduced as the bimodal random field shown in Fig.\ref{fig:model_system}. In the upper panels this field distribution is used as heterogeneous density $\rho$ in the elastic vibration equation $-\omega^2 u = \frac{1}{\rho} \nabla\cdot (G \nabla u)$, which leads to high frequency localization modes. While in the lower panel it represents the random potential $V$ in the Schr\"{o}dinger equation $\left( V - \frac{1}{2}\nabla^2\right) \psi = E\psi$, which leads to exponentially decaying soft modes at low energies.
    }
    \label{fig:eigenvectors}
\end{figure*}

Most of the phenomena important for engineering applications are described by elliptic differential operators. Through our analysis, we showed that the presence of disorder in one of the simplest mathematical operators can produce non-trivial results, such as low frequency soft modes, and high frequency Anderson-like localization. In the present section, we briefly review the similarities and differences on the localization phenomenon for differential operators used in solid mechanics, quantum mechanics and optics. For reference, Fig.~\ref{fig:eigenvectors}(a) shows typical localized modes produced by the scalar elastic vibration equation for the realization shown in Fig.~\ref{fig:model_system}.

The wavefunction $\psi$ of a single particle with energy $E$ is described by the Schr\"{o}dinger equation
\begin{equation}
    \left( V(\bm{x}) - \frac{1}{2}\nabla^2 \right) \psi = E \psi
    \label{eqn:eigen-schrodinger}
\end{equation}
where $V(\bm{x})$ is the spatially varying potential. For the ease of notation, we set $\hbar=m=1$. When the potential is periodic in space, e.g. that of a perfect crystal, space symmetry leads to the well-known extended Bloch wavefunctions~\cite{kittel1996introduction,ashcroft2010solid}. However, for random $V(\bm{x})$, there are also bound states that have very low energy and the particle gets trapped inside one of the local minima or meta-basins of $V$. Some bound states examples of the Sch\"{o}dinger equation are shown in Fig.\ref{fig:eigenvectors} (b) in contrast to those high frequency localized modes from the scalar wave equation shown in Fig.\ref{fig:eigenvectors} (a). This is a similar scenario of the soft modes as in Sect.\ref{sec:soft modes}, although here the disorder is decoupled from any derivatives of the wave function.

As a second example, we discuss the connection between Eq.~\ref{eqn:simplified-wave-1d} and electromagnetic waves. For electromagnetic problems absent of free charge and current flow, the macroscopic Maxwell's equations lead to an eigenvalue equation either for the magnetic $\bm{H}$ or the electric $\bm{E}$ fields at frequency $\omega$~\cite{joannopoulos2007photonic}
\begin{subequations}
\label{eq:EM-both}
\begin{equation}
\label{eqn:EM-master-H}
    \left( \frac{1}{\mu(\bm{x})}\nabla\times\left( \frac{1}{\epsilon(\bm{x})}\nabla\times \right) \right) \bm{H} = \frac{1}{c^2} \omega^2 \bm{H}
\end{equation}
or
\begin{equation}
\label{eqn:EM-master-E}
    \left( \frac{1}{\epsilon(\bm{x})}\nabla\times\left( \frac{1}{\mu(\bm{x})}\nabla\times \right) \right) \bm{E} = \frac{1}{c^2} \omega^2 \bm{E}
\end{equation}
\end{subequations}
where $c=1/\sqrt{\epsilon_0\mu_0}$ is the speed of light in vacuum, and $\epsilon(\bm{x})$ and $\mu(\bm{x})$ the relative electric permittivity and magnetic permeability of the medium.
For transverse magnetic (TM) modes in quasi-1D wave guides with a heterogeneous dielectric constant, the governing equation for the magnetic field $\bm{H}=\bm{e_z} H$ is
\begin{equation}
\label{eqn:waveguide-2D-epsilon-Hz}
    \nabla\cdot\frac{1}{\epsilon}\nabla H  = -\frac{1}{c^2} \omega^2 H
\end{equation}
When we assume $H=H_z(x,y)$ and $\epsilon=\epsilon(x,y)$, other terms from the expansion of the differential operator vanish and we recover the scalar elastic vibration equation, Eq.~\ref{eqn:simplified-wave-1d}. Thus, when $\epsilon$ is disordered, the magnetic field $H_z$ becomes localized, similar to the case of $G$. Given the electro-magnetic duality of Maxwell's equations, when $H_z$ is localized, $\bm{E}$ is also localized. Therefore, the high frequency localization induced by the disorder in $\rho$ and in G from Eq.\ref{eqn:simplified-wave-1d} are dual of each other. 

Recently, Filoche, Mayboroda and co-workders developed a computationally efficient method to identify localized stationary waves in finite domains~\cite{filoche2012universal,lyra2015dual,chalopin2019universality,li2017localization,filoche2017localization,piccardo2017localization}. More specifically, their technique is based on the construction of the localization landscape, which is obtained by solving the Dirichlet problem $\mathcal{L}u = 1$ with $u=0$ at the boundaries for a linear operator $\mathcal{L}$. By applying their method in both cases shown in Secs.~\ref{sec:anderson-like modes} \& \ref{sec:soft modes}, we find that the localization landscape method can only identify the existence of low frequency localized modes, namely soft modes, but not the high frequency ones with similar IPR values. We believe that our conclusion holds for any linear operator $\mathcal{L}$, and the localization landscape method is in general useful for finding soft modes.

\subsection{Connections to the Anderson model\label{sec:connection to anderson}}

The original analysis of Anderson was done for a discrete lattice model, with random on-site potential for a single particle
\begin{equation}
    H = W\sum_i V_i \ket{i}\bra{i} - t\sum_i \ket{i}\bra{i+1}
\end{equation}
where $W$ is the scale of on-site potentials, $V_i$ a random variable, and $t$ the hopping parameter. The conclusion is that the wave function of the tight-binding model will decay exponentially due to coherent scattering with the random potentials.

One observation is that the discrete Anderson model cannot be immediately mapped onto the continuously differentiable operators as studied above. More specifically, one can take the limit of small lattice spacing $\Delta x$ on a cubic lattice, which leads to the continuum form
\begin{equation}
    H = W V(x) - t( \Delta x^2 \laplacian + \nu)
\end{equation}
where by convention $t=1$. 
The single particle eigenstate satisfies the Schr\"{o}dinger equation
\begin{equation}
    \frac{\left( E + \nu - W V(x) \right) }{\Delta x^2}u + \laplacian u = 0
    \label{eqn:schrodinger-anderson}
\end{equation}
compared to the vibration equation
\begin{equation}
    \rho(x) \omega^2 u + \laplacian u = 0
    \label{eqn:vibration}
\end{equation}
Thus, every solution of $u$ in Eq.~\ref{eqn:vibration} can be regarded as a solution to Eq.~\ref{eqn:schrodinger-anderson}, with the following correspondence:
\begin{equation}
\begin{split}
    W & = 2(\rho_2-\rho_1) \omega^2 \Delta x \\ 
    V(x) & = \frac{\bar{\rho} - \rho(x)}{(\rho_2 - \rho_1)} \Delta x \\
    E & = 2\bar{\rho} \omega^2 \Delta x^2 - \nu
\end{split}
\end{equation}
where $0<\rho_1\le \rho(x) \le \rho_2$, and $\rho(x)$ follows a symmetric bimodal distribution between $\rho_1$ and $\rho_2$. This way the random potential $V(x)$ is normalized between $-1/2$ and $1/2$, and the amplitude is scaled by $W$. 
Now it is clear that scaling the random potentials $V_i$ will lead to a non-differentiable limit as $\Delta x\rightarrow 0$, invoking a stochastic Schr\"{o}dinger equation with Gaussian white noise potential~\cite{thouless1974electrons,fukushima1977spectra}. To our knowledge, in the mathematics literature, proofs for the spectrum of the continuously differentiable Schr\"{o}dinger equation have focused on the low energy band edge only, corresponding to the soft mode regime discussed in Sect.\ref{sec:soft modes}, while no proof for complete localization of the entire spectrum (hence no genuine Anderson localization) exists.
However, the tendency of more localization with increasing frequency as shown in Sect.\ref{sec:anderson-like modes} is qualitatively consistent with the Anderson model, and hence we have referred to it as Anderson-like modes.

\subsection{Extensions and applications}
In the literature of disordered solids, the Ioffe-Regel criterion~\cite{ioffe1960non}, which compares the mean free path and wavelength of waves, provides some physical understanding on the relation between high frequency localizations and cross-over from propagation to diffusion~\cite{seyf2016method}. We speculate its theoretical connections to the Coupled-Mode theory~\cite{fan2003temporal,suh2004temporal,alpeggiani2017quasinormal} (CMT) that is still advancing to this date. Future studies on multi-modal CMT will provide another perspective on the localization behaviors in domains with open boundaries, resolving the subtlety of differences and similarities between finite and infinite domains. Also, it will be interesting to investigate the applicability of these theories or their variants to signal transmission in granular matter~\cite{pinson2016signal}. Another natural conjugate problem to this study is the stress localization during material yielding, which indicates failure modes.

Other linear operators are known to display soft mode localization properties, which can also be characterized by the localization landscape~\cite{filoche2012universal}. One example is the bi-harmonic operator that is relevant to elastic buckling problems~\cite{audoly2020localization,groh2019role} when heterogeneities in the elastic modulus and density are present.

Our results provide insights to metamaterial design and optimization for given target modes/functionalities. There are several examples where optimizing the structure of the material to produce localized modes at specific vibrational frequencies can help build systems with specific response and desired properties, such as rectification, flexibility, stiffness, etc.~\cite{boechler2011bifurcation,matlack2018designing,cha2018electrical,cai2019unbounded,fang2014acoustic, kim2020multifunctional,marthelot2018designing}. For example, one can use the studied framework to properly choose the complementary material properties (for example density v.s. modulus on the Ashby plot~\cite{ashby1993materials}) of the components that gives desired localization or mode selection. 

\section{Summary}
In this paper, we study the effects of disordered $\rho$ and $G$ for the scalar wave equation. We focus our analysis on finite domains of length $L$ with bimodal disorder with spatial correlation $l_c$. When either $G$ or $\rho$ is disordered, we show that at high frequencies Anderson-like localized modes appear near the domain boundaries, and are characterized by large values of inverse participation ratio. We also show the emergence of low frequency soft modes when the modulus $G$ is heterogeneous and correlation length is large enough. 
We demonstrate the predictions of our statistical analysis, by studying the propagation of mechanical waves in a medium with disordered $G$ under external forced vibrations. We find that by increasing the excitation frequencies, there is a cross-over from propagating waves to localized ones, reminiscent of the Ioffe-Regel frequency in amorphous solids models. Finally, we discuss the similarities and differences between linear differential operators similar to the scalar wave equation, and show the specific conditions for which the Maxwell's equations result in identical predictions to our study.

\section*{Acknowledgement}
We are grateful to A. Tanguy, S. Yip, S.G. Johnson, Z. He, M. Li, N. Nadkarni, M. Mirzadeh, M. Benzaouia and P. Lee for insightful discussions. 
All authors contributed equally to this work.

\section*{Appendix}

\subsection*{Numerical Method for eigenvalue and Dirichlet problems}
In the present section, we describe the numerical procedure that is followed to solve the equations discussed previously. For simplicity, we will focus on a scalar field $u$ as the unknown. The equation we are interested in is
\begin{equation}
    \rho\left(\bm{x}\right)u - \nabla\cdot\left(G\left(\bm{x}\right)\nabla u \right) = 0
    \label{eq:fem_eq}
\end{equation}

\subsubsection{Finite Element Formulation}
We use the finite element method (FEM) for the discretization of eq.~\ref{eq:fem_eq} although one can use another method (e.g. finite differences, pseudo-spectral, spectral elements) for approximating the solution. The common procedure in FEM discretization is first to define a mesh over the domain of interest. This mesh consists of a collection of nodes that are connected to form elements. Element shapes that are commonly used in practice for arbitrary geometries are triangles, in two-dimensions, and tetrahedrons, in three-dimensions. Other shapes can be used as well, e.g. rectangles or hexahedrons, but the mesh generation in arbitrary shaped domains then becomes a laborious task. 

The second step in FEM is the choice of basis functions that are used to interpret the unknown field $u$ over the discretized domain. In general, the basis function which corresponds to the unknown $i$, $\phi_i(\bm{x})$, is chosen to be local in space, meaning that it is non-zero only within the elements where $i$ belongs in.

The third step is to turn eq.~\ref{eq:fem_eq} in the FEM formulation. To do so, we multiply the equation with $\phi_i(\bm{x})$ and integrate it over the domain of interest
\begin{equation}
    \int\left[\rho\left(\bm{x}\right)u - \nabla\cdot\left(G\left(\bm{x}\right)\nabla u \right)\right]\phi_i\left(\bm{x}\right)\,dV = \int f\left(\bm{x}\right)u \phi_i\left(\bm{x}\right)\,dV
    \label{eq:fem_eq_integrated}
\end{equation}
A common step is to reduce higher order derivatives by performing the divergence theorem as follows  
\begin{equation}
\begin{split}
    \int \nabla\cdot\left(G\left(\bm{x}\right)\nabla u \right)\phi_i\left(\bm{x}\right)\,dV = & - \int \left(G\left(\bm{x}\right)\nabla u \right) \cdot \nabla\phi_i\left(\bm{x}\right)\,dV \\
    & + \int \bm{n}\cdot \left(\phi_i\left(\bm{x}\right) G\left(\bm{x}\right)\right)\nabla u\,dA
    \label{eq:fem_div_thrm}
\end{split}
\end{equation}
where a surface integral term appears. Divergence theorem is useful for two reasons: i) the first is lowering the highest order derivative, which allows for the use of lower order polynomials in approximating the unknown field $u$; ii) and the second is the natural implementation of Neumann and Robin boundary conditions. The integrals in eq.~\ref{eq:fem_eq_integrated} can be written as the summation of individual integrals over each element generated through the step of mesh generation, leading to
\begin{equation}
\begin{split}
    &\sum_j \left[\int \rho\left(\bm{x}\right)u\phi_i\left(\bm{x}\right) + \left(G\left(\bm{x}\right)\nabla u \right) \cdot \nabla\phi_i\left(\bm{x}\right) \right]\,dV_j = \\
    &\sum_j\left[\int f\left(\bm{x}\right)u \phi_i\left(\bm{x}\right)\,dV_j + \int \bm{n}\cdot \left(\phi_i\left(\bm{x}\right) G\left(\bm{x}\right)\right)\nabla u\,dA_j \right]
    \label{eq:fem_eq_integrated}
\end{split}
\end{equation}

The final step before generating the set of algebraic equations from eq.~\ref{eq:fem_eq} is to define the expansion for the unknown field $u$. Herein, we follow Galerkin approach, which considers the expansion polynomials of $u$ to be the same as the projection basis function $\phi_i$ considered in the previous equations. Therefore, we can write $u$ as
\begin{equation}
    u = \sum_k u_k \phi_k\left(\bm{x}\right)
    \label{eq:exp_u_fem}
\end{equation}
where $u_k$ are the values of $u$ at the nodes of the generated mesh. Substituting eq.~\ref{eq:exp_u_fem} in the FEM formulation we arrive at the final step before we generating the system for the algebraic unknowns, $\bm{u}=\{u_k\}$
\begin{equation}
\begin{split}
    &\sum_k \sum_j \left[\int \rho u_k\phi_k\phi_i + G u_k\nabla \phi_k \cdot \nabla\phi_i \right]\,dV_j = \\
    &\sum_k \sum_j\left[\int f u_k \phi_k \phi_i\,dV_j + \int \bm{n}\cdot \phi_i G u_k\nabla \phi_k\, dA_j \right]
    \label{eq:fem_eq_integrated_galerkin}
\end{split}
\end{equation}
where for simplicity we dropped all the spatial dependencies from all functions. The final step is to perform the integration and form the system of algebraic unknowns. The most natural way to approximate the integrals is through quadrature. Here, we use the Gauss-Legendre rule leading to 
\begin{equation}
\begin{split}
    &\sum_{q_V} \sum_k \sum_j \left[ \rho u_k\phi_k\phi_i + G u_k\nabla \phi_k \cdot \nabla\phi_i \right]\,w_{q_V}J_{q_V} = \\
    &\sum_k \sum_j\left[\sum_{q_V} f u_k \phi_k \phi_i\,w_{q_V}J_{q_V} +  \sum_{q_A}\bm{n}\cdot \phi_i G u_k\nabla \phi_k\, w_{q_A}J_{q_A}\right] 
    \label{eq:fem_eq_quadrature}
\end{split}
\end{equation}
where $w_{q_{V/A}}$ and $J_{q_{V/A}}$ are the integration weights and the Jacobian of transformation between the physical and the Gauss-Legendre quadrature space, respectively, for the volume and surface integrals. 

Eq.~\ref{eq:fem_eq_quadrature} is in the final form to set our algebraic system of equations
\begin{equation}
\begin{split}
    \bm{A}\bm{u} = \bm{b}
    \label{eq:fem_eq_discrete_sys}
\end{split}
\end{equation}
where $\bm{b}$ is the right hand side (RHS) of the above equations. The solution of $\bm{u}$ can be found by inverting the matrix $\bm{A}$. There are several methods to do that which either rely on direct inversion of the matrix (e.g. LU decomposition) or on iterative approximations which are based on a minimization problem (e.g. Generalized Minimal Residual).

\subsubsection{Eigenvalue Problem}
In the cases we are interested for the eigenvalues of the operator
$$
\mathcal{L}=-\frac{1}{\rho}\nabla\cdot\left(G\nabla \right)
$$ 
we can formulate eq.~\ref{eq:fem_eq_quadrature} as the discrete eigenvalue of eq.~\ref{eq:fem_eq}. The original eigenvalue problem is expressed as
\begin{equation}
    \mathcal{L}u=\lambda u
    \label{eq:eignvl_bfem}
\end{equation}
where $\lambda$ corresponds to the eigenvalue of the operator. Performing exactly the same procedure for discretizing the operator using FEM, we arrive at the following form of eq.~\ref{eq:eignvl_bfem}
\begin{equation}
\begin{split}
    &\sum_k \sum_j \left[\sum_{q_V} \left( G u_k\nabla \phi_k \cdot \nabla\left(\frac{\phi_i}{\rho}\right) \right)\,w_{q_V}J_{q_V} - \right. \\
    & \left. \sum_{q_A}\bm{n}\cdot \phi_i \left(\frac{G}{\rho}\right) u_k\nabla \phi_k\, w_{q_A}J_{q_A} \right] = \\
    & \lambda \sum_{q_V} \sum_k \sum_j  u_k \phi_k \phi_i\,w_{q_V}J_{q_V} 
    \label{eq:fem_eq_quadrature_eigen}
\end{split}
\end{equation}
which can be written in a more compact form
\begin{equation}
\begin{split}
    \bm{A}\bm{u} = \lambda \bm{M}\bm{u}
    \label{eq:fem_eq_discrete_sys_eigen}
\end{split}
\end{equation}
where $\bm{M}$ is also known as mass matrix. The common procedure to numerically find the eigenvalues of eq.~\ref{eq:fem_eq_discrete_sys_eigen} is to transform it as follows
\begin{equation}
\begin{split}
    \bm{M}^{-1}\bm{A}\bm{u} = \lambda\bm{u}
    \label{eq:fem_eq_discrete_sys_eigen_frndl}
\end{split}
\end{equation}
with which iterative eigenvalue methods for large-scale systems (e.g. Power, Lanczos, Arnoldi) work the best.

For some boundary conditions, e.g. Dirichlet, the matrix $\bm{M}$ can become singular, and therefore we cannot formulate the problem in the form of eq.~\ref{eq:fem_eq_discrete_sys_eigen_frndl}. To overcome this difficult, we solve a modified eigenvalue problem that shares the same eigenvalues as the original one. The most common method to use is the shift-invert spectral transformation. More specifically, we transform eq.~\ref{eq:fem_eq_discrete_sys_eigen} in the following form
\begin{equation}
\begin{split}
    \left(\bm{A}-\sigma\bm{M}\right)^{-1}\bm{M}
    \bm{u} = \nu\bm{u}
    \label{eq:shift_invert}
\end{split}
\end{equation}
where $\sigma$ is the shift in the eigenvalue spectrum we perform, and $\nu$ is the eigenvalue of the modified problem. Based on the commutative nature of the operators, the original eigenvalue $\lambda$ is connected with the $\nu$ through the following algebraic relation
\begin{equation}
\begin{split}
    \lambda = \sigma + 1/\nu
    \label{eq:lambda_nu}
\end{split}
\end{equation}
Eq.~\ref{eq:shift_invert} involves the inversion of a matrix which can be easily performed using a linear solver, similar to those discussed in the previous section. In the present study, we use ARPACK library that implements Implicitly Restarted Arnoldi Method to find the eigenvalue spectrum of the operators involved herein. 

\subsection*{Dynamical simulations}
\begin{figure}
    \centering
    \includegraphics[width=\linewidth]{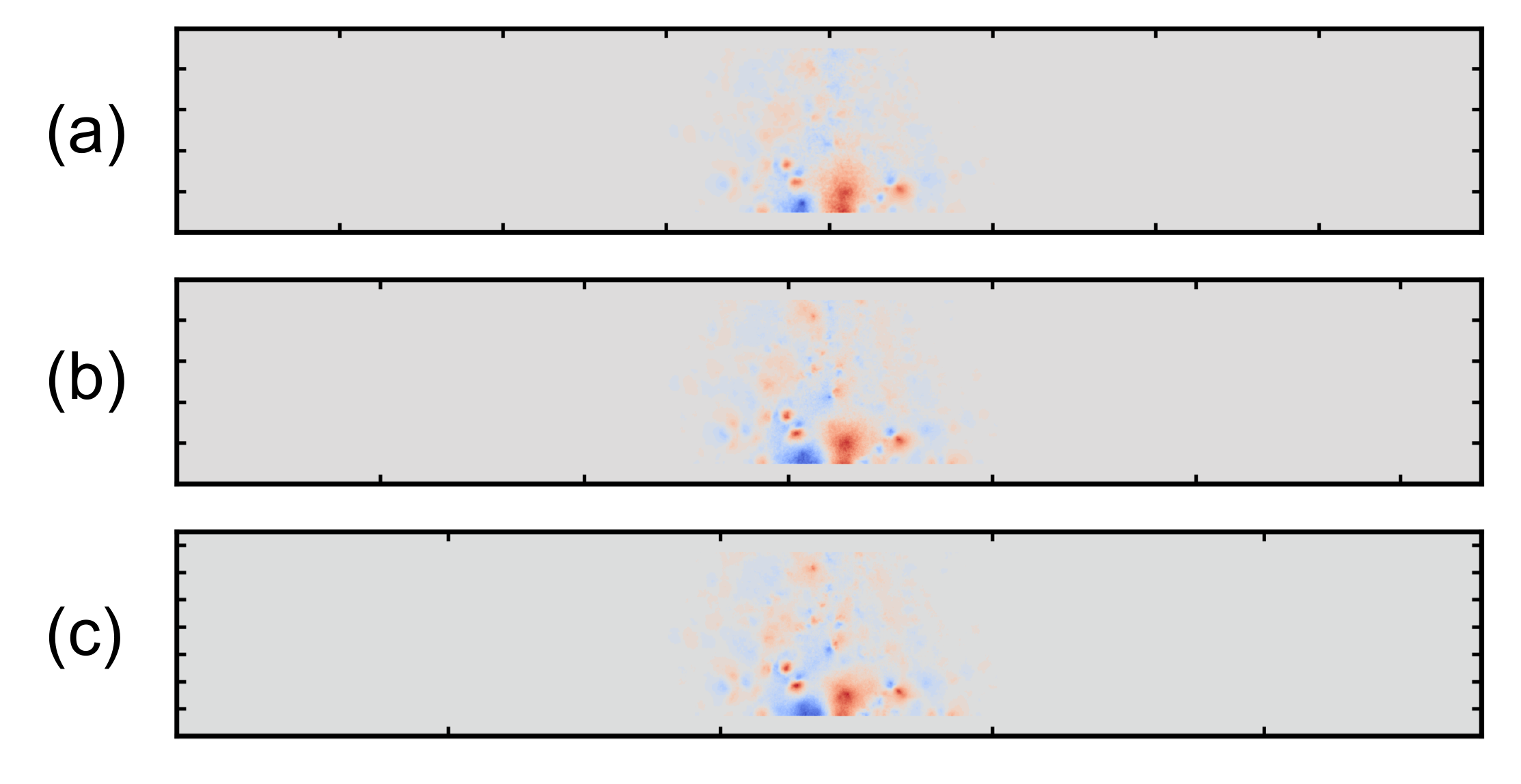}
    \caption{Convergence of MEEP simulations. (a-c) showing the $E_x$ field at the same stopping time for pixel resolution (definition see ~\cite{oskooi2010meep}) 10, 20 and 30. Color range for all three resolutions is fixed the same.}
    \label{fig:appendix-diffuson-convergence}
\end{figure}

For the simulations in Sect.\ref{sec:diffuson}, we take advantage of the mathematical equivalence between the elastic wave (Eq.\ref{eqn:simplified-wave-1d}) and the quasi-1D wave guide (Eq.\ref{eqn:waveguide-2D-epsilon-Hz}) as described in Sect.\ref{sec:nature of localization-EM}. Here the free boundary condition is realized through the Perfect-Magnetic-Conductor (PMC) layers~\cite{joannopoulos2007photonic,lindell2016electromagnetic}, which essentially enforces H$_x$=H$_y$=0 at $z=0$ and $z=L_z$. The open (radiating) boundaries along the wave guide x direction is enforced by adding Perfectly-Matched-Layers (PML)~\cite{johnson2008notes,berenger1994perfectly} at the ends. A pulse of continuous line source of frequency $\omega$ is initiated in the middle of the domain along the z-axis and lasts for a reasonable long time. Then we observe the H$_z$ field profile at later times for different values of $\omega$. 
The corresponding E field is shown for its x component in Fig.\ref{fig:appendix-diffuson-convergence}. The external driving is done through either H$_z$ field or the E fields and as expected which field component to excite does not affect the results.
The MEEP interface scripts are available from the authors upon requests. 

\subsection*{Solutions for cylindrical symmetry geometry}
We solve the toy model as in Fig.3 in a polar coordinate. The material properties are defined as:
$\rho=\rho_1$, $G=G_1$ for $0<r<R_{in}$; 
$\rho=\rho_2$, $G=G_2$ for $R_{in}<r<R_{out}$. The eigenfunction for eigenvalue $\omega^2$ can be written as a piece-wise function 
\begin{equation}
    u_i = f_i(r) e^{i m\theta} \quad i=0,1
\end{equation}
where the radial parts are 
\begin{equation}
\begin{cases}
f_0 & = J_m (k_0 r) \quad 0<r<R_{in} \\
f_1 & = A J_m (k_1 r) + B N_m (k_1 r) \quad R_{in} < r < R_{out}
\end{cases}
\end{equation}
Here 
\begin{equation}
    k_i = \omega \sqrt{\frac{\rho_i}{G_i}}
\end{equation}
is the wavenumber, $J_m$ and $N_m$ the m-th Bessel function and Neumann function.
The boundary conditions are
\begin{equation}
\begin{cases}
    G_1 \frac{d}{dr}f_0(R_{in}) & = G_2 \frac{d}{dr}f_1(R_{in}) \\
    f_1 (R_{out}) & = 0
\end{cases}
\end{equation}
Mathematica scripts for evaluating the eigenvalues of this set of Bessel functions are available upon reasonable request to the authors.

\bibliography{localization}

\end{document}